\documentclass[twocolumn, twocolappendix]{aastex631}
\usepackage{comment}
\usepackage{amsmath}
\usepackage{graphicx}
\usepackage[caption=false]{subfig}
\begin{document}

\title{Another piece to the puzzle: radio detection of a JWST detected AGN candidate}

\author[0009-0009-8274-441X]{Anniek J. Gloudemans}
\affiliation{NSF NOIRLab, Gemini Observatory, 
670 N A'ohoku Place
Hilo, HI 96720, USA; anniek.gloudemans@noirlab.edu}

\author[0000-0001-6889-8388]{Kenneth J. Duncan}
\affiliation{Institute for Astronomy, Royal Observatory, Blackford Hill, Edinburgh, EH9 3HJ, UK}

\author[0000-0003-2895-6218]{Anna-Christina Eilers}
\affiliation{MIT Kavli Institute for Astrophysics and Space Research, 77 Massachusetts Avenue, Cambridge, 02139, Massachusetts, USA}
\affiliation{Department of Physics, Massachusetts Institute of Technology, Cambridge, MA 02139, USA}

\author[0000-0002-6822-2254]{Emanuele Paolo Farina}
\affiliation{International Gemini Observatory/NSF NOIRLab, 670 N A’ohoku Place, Hilo, Hawai'i 96720, USA}

\author[0000-0002-6047-430X]{Yuichi Harikane}
\affiliation{Institute for Cosmic Ray Research, the University of Tokyo, 5-1-5 Kashiwa-no-Ha, Kashiwa City, Chiba, 277-8582, Japan}

\author[0000-0001-9840-4959]{Kohei Inayoshi}
\affiliation{Kavli Institute for Astronomy and Astrophysics, Peking University, Beijing 100871, China}

\author[0000-0003-3216-7190]{Erini Lambrides}
\affiliation{NASA Goddard Space Flight Center, Code 662, Greenbelt, 20771, MD, USA}

\author[0000-0002-4437-1773]{Eleni Vardoulaki}
\affiliation{IAASARS, National Observatory Athens, Lofos Nymfon, 11852 Athens, Greece}

\begin{abstract} 
Radio observations can provide crucial insight into the nature of a new abundant and mysterious population of dust-reddened active galactic nuclei (AGN) candidates discovered by the James Webb Space Telescope (JWST), including ``Little Red Dots" (LRDs). In this study, we search for radio bright sources in a large sample of $\sim$700 JWST discovered AGN candidates ($z\sim2-11$) in the $0.144–3$ GHz frequency range, utilizing deep radio imaging in COSMOS, GOODS-N, and GOODS-S. Only one source, PRIMER-COS 3866 at $z=4.66$, is significantly detected in our radio surveys, which has been previously identified as an X-ray AGN. Its radio properties are consistent with both an AGN and star formation origin with a spectral index of $\alpha=-0.76^{+0.11}_{-0.09}$, radio-loudness of $R\approx0.5$, and brightness temperature limit of $T_b \gtrsim 10^{3}$ K. Our stacking results of both spectroscopically and photometrically selected AGN candidates yield non-detections in all fields, with 3$\sigma$ limits of $L_{1.4\text{GHz}} < 8.6\times10^{39}$ erg s$^{-1}$ (spectroscopic sample) and $L_{1.3\text{GHz}} < 1.3\times10^{39}$ erg s$^{-1}$ (photometric sample). We demonstrate that these results are still consistent with expectations from the empirical $L_X - L_{\text{H}\alpha}$ and $L_X - L_R$ correlations established for local AGN. We argue that current radio observations in these studied fields have insufficient depth to claim JWST discovered AGN candidates are radio-weak. We project that future surveys carried out by the SKA and ngVLA should be able to obtain significant detections within a few hours, providing crucial measurements of their brightness temperature, which would allow for distinguishing between AGN and starburst-driven origins of this new abundant population.

\end{abstract}


\section{Introduction} \label{sec:intro} 

The first years of JWST observations have led to the discovery of an abundant population of faint high-$z$ AGN candidates. This includes the so-called ``Little Red Dots'': a puzzling population of faint sources at high-$z$ with broad emission lines, compact morphology, and extremely red color (LRDs; \citealt{ Kocevski2023ApJ...954L...4K, Harikane2023ApJ...959...39H, Labbe2023arXiv230607320L, Matthee2024ApJ...963..129M}). Surprisingly, some studies suggest that the number densities of these AGN candidates are more than an order of magnitude higher than expected from extrapolating quasar luminosity functions (e.g. \citealt{Kokorev2024ApJ...968...38K, Pizzati2024arXiv240918208P, Kocevski2024arXiv240403576K}) and their supermassive black hole (SMBH) masses are high compared to estimated host galaxy masses, suggesting extreme black hole growth (e.g. \citealt{Kokorev2023ApJ...957L...7K, Maiolino2023arXiv230801230M, Greene2024ApJ...964...39G}). In addition, a lack of hot torus emission expected from AGN has been reported (e.g. \citealt{Williams2024ApJ...968...34W, Wang2024arXiv240302304W}). Furthermore, \cite{Inayoshi2024ApJ...973L..49I} argue that their radiative efficiency is expected to be as high as 20\%, which would imply high BH spin and ideal conditions for radio jet formation. These sources might therefore prove to be a crucial phase in early black hole and galaxy formation.

Alternatively, it has been argued that the newly discovered sources could instead be compact massive galaxies with extreme starburst activity (e.g. \citealt{Baggen2024arXiv240807745B, Perez-Gonzalez2024ApJ...968....4P}) and ionized outflows have been detected which could play a role in emission line broadening (e.g. \citealt{Wang2024arXiv240302304W}). 
To investigate different physical processes this new population has been studied at many different wavelengths. A recent study by \cite{Yue2024ApJ...974L..26Y} obtained a tentative detection at X-ray wavelengths, further strengthening the case for their AGN nature. However, their X-ray emission is $\sim$1 dex lower than expected from the $L_X - L_{\text{H}\alpha}$ relation, which suggests that these are different from previously observed type-I AGN (see also \citealt{Maiolino2024arXiv240500504M}). \cite{Lambrides2024arXiv240913047L} found in addition to significantly weaker than expected X-ray emission, a lack of high-ionization rest-UV lines in a sample JWST broad line AGN at $z>4$. In the ALMA sub-mm regime, none of these sources have been detected individually nor in a stack, providing strong limits on the contribution of obscured star formation \citep{Labbe2023arXiv230607320L, Akins2024arXiv240610341A}.

Radio observations are a powerful tool to help determine their nature. Both star formation and AGN can contribute to synchrotron radio emission (see e.g. \citealt{best2023arXiv230505782B}), however, properties such as radio-loudness (e.g. \citealt{Macfarlane2021MNRAS.506.5888M}), brightness temperature (e.g. \citealt{Morabito2022MNRAS.515.5758M}), and the radio spectral index (e.g. \citealt{Randall2012MNRAS.421.1644R}) can provide crucial evidence to distinguish between an AGN or stellar origin. Recent work has investigated the radio properties of JWST discovered AGN candidates (e.g. \citealt{Akins2024arXiv240610341A, Mazzolari2024arXiv241204224M, Perger2024arXiv241119518P}). However, all of them reported non-detections from both individual sources and image stacking analysis. In this work, we extend this search to multiple well-studied fields and multiple radio facilities including the Low Frequency Array (LOFAR; \citealt{vanHaarlem2013A&A...556A...2V}), MeerKAT \citep{Jonas2016mks..confE...1J}, and the Very Large Array (VLA; \citealt{Thompson1980ApJS...44..151T}). Furthermore, we present the expected potential detections by future radio surveys. Throughout this paper, we assume a $\Lambda$-CDM cosmology with $H_{0}$= 70 km s$^{-1}$ Mpc$^{-1}$, $\Omega_{M}$ = 0.3, and $\Omega_{\Lambda}$ = 0.7.

\section{Data}

\begin{table}
    \centering
    \caption{Candidate AGN samples}
    \label{tab:tab_lrds}
    \begin{tabular}{c l c}
        \hline \hline
        COSMOS & $10^{\text{h}}00^{\text{m}}27.40^{\text{s}} +2^{\circ}13' 50''.64$ & \\ 
        \hline \hline
        & \cite{Kocevski2024arXiv240403576K} & 81$^{*}$ \\
        & \cite{Akins2024arXiv240610341A} & 434$^{*}$ \\
        & \cite{Kokorev2024ApJ...968...38K} & 78$^{*}$ \\
        \hline \hline
        GOODS-N & $12^{\text{h}}36^{\text{m}}49.4^{\text{s}} +62^{\circ} 12' 58''$ & \\
        \hline \hline
        & \cite{Matthee2024ApJ...963..129M} & 7 \\
        & \cite{Maiolino2023arXiv230801230M} & 10 \\  
        \hline \hline
        GOODS-S & $03^{\text{h}}32^{\text{m}}28^{\text{s}} -27^{\circ} 48' 30''$ & \\
        \hline \hline
        & \cite{Matthee2024ApJ...963..129M} & 1 \\
        & \cite{Maiolino2023arXiv230801230M} & 2 \\
        & \cite{Kocevski2024arXiv240403576K} & 56$^{*}$ \\
        & \cite{Kokorev2024ApJ...968...38K} & 27$^{*}$ \\
        \hline \vspace{0.1pt}
    \end{tabular} \\
    \vspace{0.0cm} \raggedright {\textbf{Notes.} $^{*}$The majority of AGN candidates in the sample are not spectroscopically confirmed }
\end{table}

\begin{table*}
    \centering
    \caption{Details on available radio observations}
    \label{tab:tab_radio_obs}
    \begin{tabular}{l l l l l l l l l }
        \hline \hline
        Field & \# sources & Telescope & Survey & Freq. & Resolution &  rms depth & $F_{\nu}^{*}$ \\
         & & & & (GHz) & (\arcsec) & ($\mu$Jy beam$^{-1}$) & ($\mu$Jy beam$^{-1}$)\\   
        \hline
        COSMOS & 593 & LOFAR & COSMOS-LOFAR & 0.144 & 6 $\times$ 12 & 150 & $<17$ \\
         & & MeerKAT & MIGHTEE DR1 & $\sim$1.3 & 8.9 & 3.5 & $<0.44$ \\ 
         & & VLA & VLA-COSMOS 3 GHz LP & 3 & 0.75 & 2.3 & $<0.41$ \\ 
         \hline
        GOODS-N & 17 & LOFAR & LoTSS-DR2 & 0.144 & 6 & 70 & $<59$ \\
         & &  VLA & GOODS-N VLA & 1.4 & 1.7 & 4 & $<3.7$ \\ 
        \hline
        GOODS-S & 86 & MeerKAT & MIGHTEE DR1 & $\sim$1.3 & 5.5 & 1.9 & $<0.80$ \\
        \hline \vspace{0.1pt}
    \end{tabular} \\
    \vspace{0.0cm} \raggedright {\small \textbf{Notes.} $^{*}$This column gives the 3$\sigma$ flux density limits derived from the stacks.}
\end{table*}

The fields studied in this work are selected based on availability of deep (public) radio data and JWST discovered AGN candidates. Based on this, we choose to study the radio properties of AGN candidates in three well-known fields: Cosmic Evolution Survey (COSMOS), Great Observatories Origins Deep Survey North (GOODS-N) and South (GOODS-S). The GOODS-S field is embedded within the Chandra Deep Field South (CDFS) field and therefore we utilize multi-wavelength data from both fields. 

\subsection{AGN candidate sample}
\label{subsec:sample}

This work focuses on five samples of AGN candidates in COSMOS, GOODS-N, and GOODS-S discovered by \cite{Maiolino2023arXiv230801230M}, \cite{Matthee2024ApJ...963..129M}, \cite{Kocevski2024arXiv240403576K}, \cite{Akins2024arXiv240610341A}, and \cite{Kokorev2024ApJ...968...38K}. A summary of the fields and samples is given in Table~\ref{tab:tab_lrds}. The sample of \cite{Maiolino2023arXiv230801230M} contains 10 sources in GOODS-N and 2 sources in GOODS-S, which have been obtained from the JADES survey \citep{Eisenstein2023arXiv230602465E} with spectroscopic redshifts ranging from 4.1-6.8. We note that only $\sim$20\% of these sources would be photometrically selected as LRD from their colors.
The samples of \cite{Matthee2024ApJ...963..129M} have been obtained from the EIGER \citep{Kashino2023ApJ...950...66K} and FRESCO \citep{Oesch2023MNRAS.525.2864O} surveys in multiple fields including GOODS-N and -S. In their work, they present the discovery of 20 new AGN candidates at $z=4.2-5.5$, including 7 in GOODS-N and 2 in GOODS-S.
The AGN candidate samples of \cite{Kocevski2024arXiv240403576K} have been obtained from the PRIMER NIRCam survey in COSMOS and the JADES and NGDEEP survey \citep{Bagley2024ApJ...965L...6B} in GOODS-S. The sample of \cite{Kocevski2024arXiv240403576K} contains 81 sources in COSMOS and 56 in GOODS-S, however, only 1 of these sources is spectroscopically confirmed in each field. The vast majority of sources in their sample therefore only have photometric redshifts, ranging from 2.4-11.9 (with a median $z_{\text{phot}}=6.1$). 
The sample of \cite{Akins2024arXiv240610341A} contains 434 candidate AGN that have been photometrically selected from the COSMOS-Web survey \citep{Casey2023ApJ...954...31C} with photometric redshift ranging from 4.5-9.4 (median $z_{\text{phot}}=$6.3).
Finally, we include 27 and 78 AGN candidates from \cite{Kokorev2024ApJ...968...38K} in the GOODS-S and COSMOS fields, respectively. These have also been photometrically selected using data from multiple JWST programs: PRIMER (in COSMOS), FRESCO, JADES, and JEMS (\citealt{Williams2023ApJS..268...64W}, in GOODS-S).

It is important to note that these photometric selected samples not only have uncertain redshifts, but also no confirmed broad Balmer emission lines. The samples studied in this work therefore likely consist of multiple populations of sources. However, to increase our chances of finding radio detections, we investigate both photometrically and spectroscopically selected AGN candidates. The results for the photometrically and spectroscopically selected samples are presented separately where possible. 

\subsection{Radio data}
\label{subsec:radio_data}

Table~\ref{tab:tab_radio_obs} presents the radio surveys used in this work, including observations from LOFAR, MeerKAT, and VLA. The LOFAR observations at 144 MHz consist of data from the LOFAR Two Metre Sky Survey Data Release 2 (LoTSS-DR2; \citealt{Shimwell2022A&A...659A...1S}) and the COSMOS-LOFAR survey (DDT19\_002, PI: Vardoulaki, Vardoulaki et al. in prep.). The LoTSS-DR2 images are obtained with 8 hours of observations per pointing with an rms depth of $\sim$70 $\mu$Jy beam$^{-1}$ in the GOODS-N field. The COSMOS map has been imaged using 48 hours of observations, however, due to the low declination the rms depth is lower than GOODS-N with $\sim$150 $\mu$Jy beam$^{-1}$.

The first data release of the MeerKAT International Gigahertz Tiered Extragalactic Explorations (MIGHTEE; \citealt{Jarvis2016mks..confE...6J}) continuum survey contains deep imaging in COSMOS and GOODS-S at $\sim$1.3 GHz (see \cite{Hale2024arXiv241104958H} for details). The radio images are provided in two resolutions for each field\footnote{See \url{https://archive-gw-1.kat.ac.za/public/repository/10.48479/7msw-r692/index.html}}. In this work, we utilize the high-resolution images (5-9\arcsec), which are not confusion limited and provide the deepest images in this particular case.

VLA data is publicly available in the COSMOS and GOODS-N fields at 3 and 1.4 GHz, respectively. The VLA-COSMOS 3 GHz Large Project \citep{smolcic2017A&A...602A...6S} reaches a depth of 2.3 $\mu$Jy beam$^{-1}$ (from 384 hours of observations) with a high resolution of 0.75\arcsec. The GOODS-N VLA map includes 165 hours of observations and reaches $\sim$4 $\mu$Jy beam$^{-1}$ noise level at 1.4 GHz with a resolution of 1.7\arcsec\ \citep{Morrison2010ApJS..188..178M}. 

All sources in our samples are covered by these radio surveys and together they cover a broad range of radio frequencies. To determine if any AGN candidates are detected in these surveys, we measure the SNR at the location of each source by dividing the peak flux density by the local rms (within 60\arcsec). 
This process yields multiple $>5\sigma$ detections in each of the fields (1 in GOODS-N, 5 in GOODS-S, 36 in COSMOS), however, upon visual inspection the vast majority of these radio sources do not seem to be associated with the AGN candidate, but with a bright galaxy in the vicinity (a few to tens of arcseconds away; see Appendix~\ref{appendix:radio_association}). We identify one source with associated radio emission in the COSMOS field, namely PRIMER-COS 3866 at $z=4.66$. We extract its radio flux density using the Python Blob Detector and Source Finder (\textsc{PyBDSF}; \citealt{Mohan2015ascl.soft02007M}). The radio properties of this source are discussed next. 

\section{Radio detection of PRIMER-COS 3866}
\label{sec:primer_cos_3866_detection}

\begin{figure*}
\centering 
\subfloat[]{%
  \includegraphics[width=0.86\columnwidth, trim={0.2cm 0cm 0cm 0.0cm}]{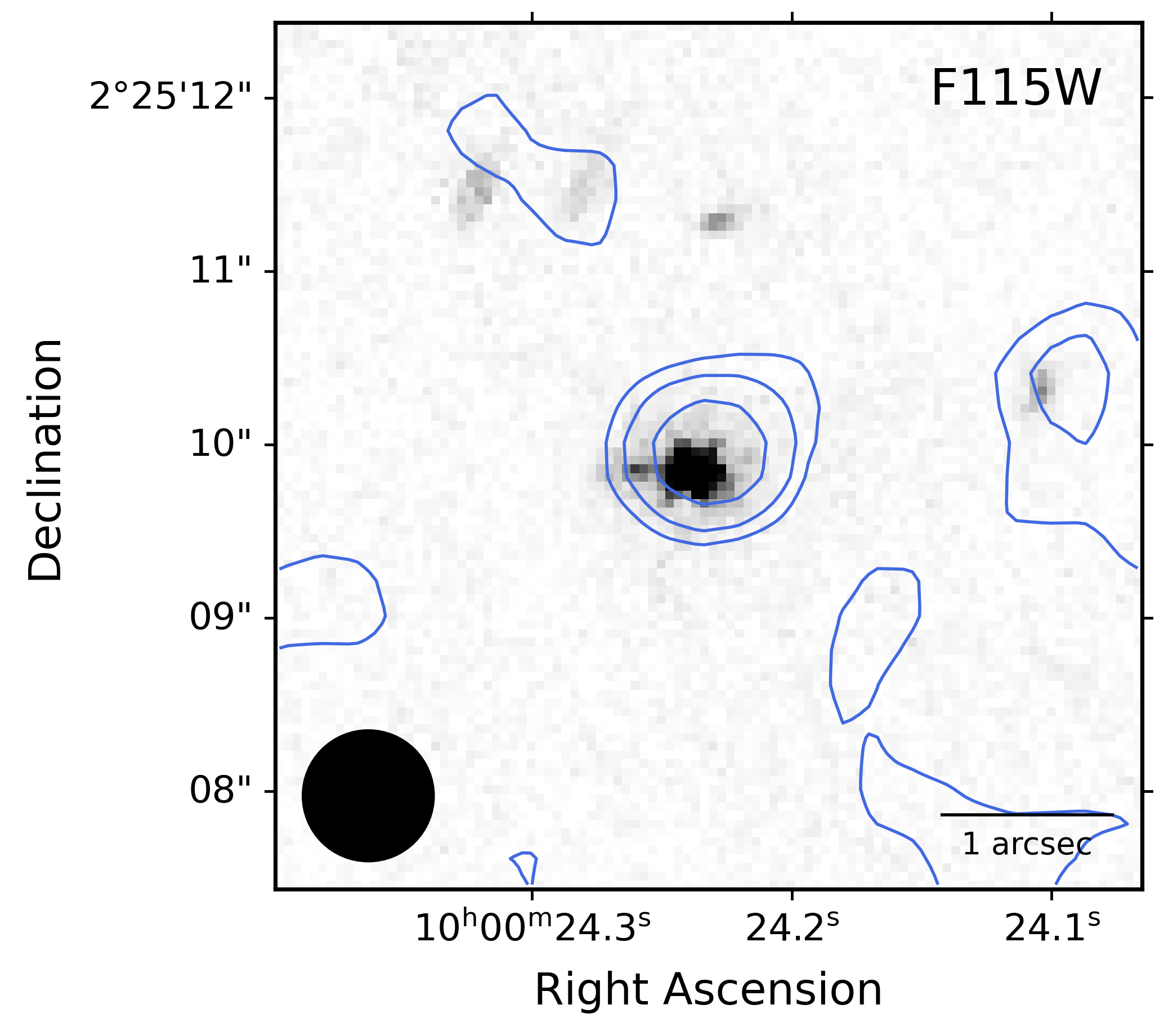}
}\qquad
\subfloat[]{%
  \includegraphics[width=1.15\columnwidth, trim={0.1cm 0cm 0cm 0.0cm}]{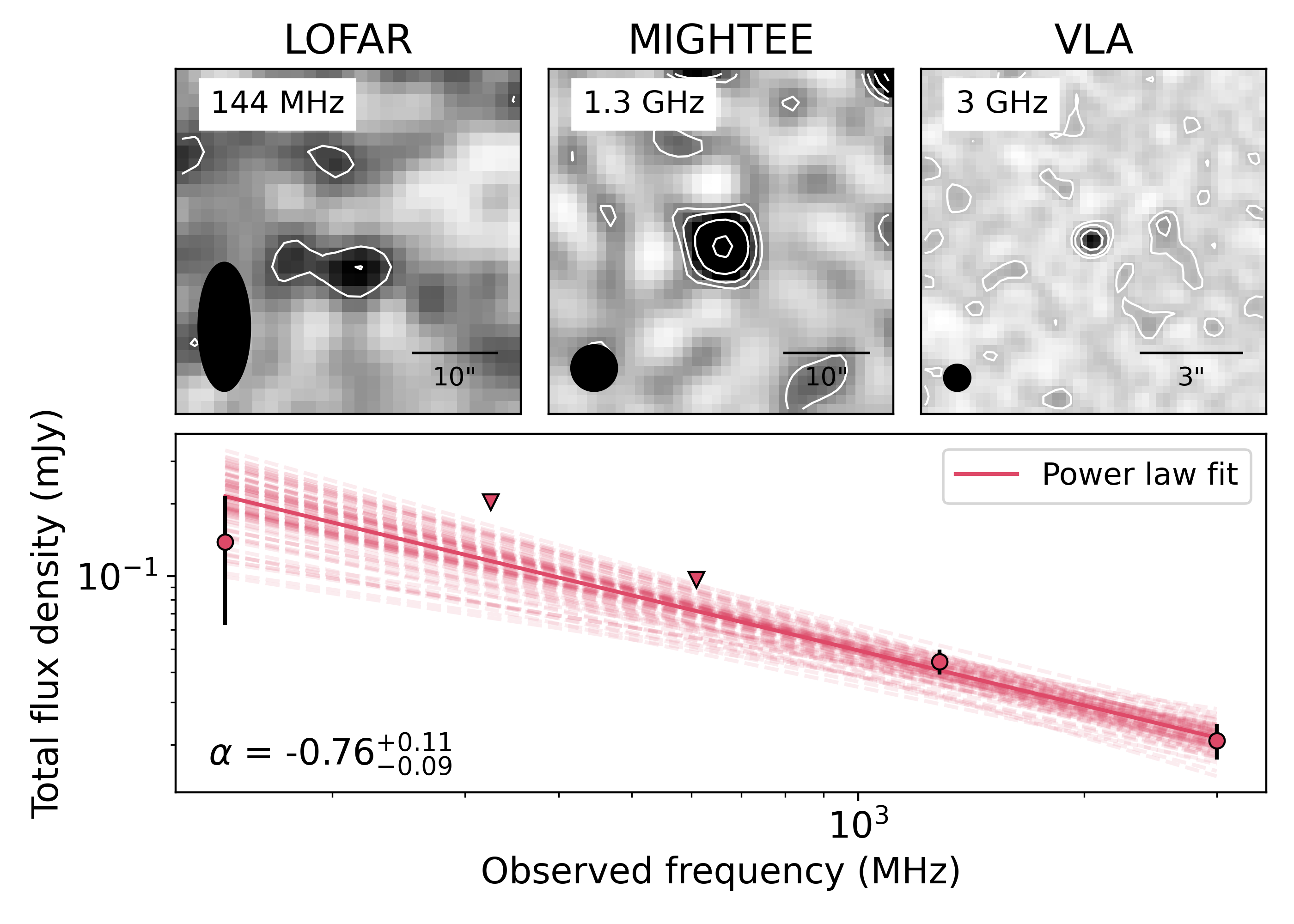}
}\vspace{-0.3cm}
\caption{\label{fig:primer_cos_3866} \textbf{Left:} F115W image of PRIMER-COS 3866 at $z=4.66$ with VLA radio contours drawn at [$-1$, 0.5, 1, 2, 4]$\ \times\ 3\sigma_{\text{rms}}$. The beam size is shown in the lower left corner. \textbf{Right:} Top: Radio cutouts images of PRIMER-COS 3866. This source is detected in LOFAR, MIGHTEE, and VLA at 144 MHz, 1.3 GHz, and 3 GHz, respectively. The beam size of the surveys and linear scale are shown in the corners of each cutout. Bottom: Radio spectrum of PRIMER-COS 3866, including 3$\sigma$ upper limits at 325 and 610 MHz from GMRT. A power law fit to the total flux densities resulted in a radio spectral index value of $\alpha=-0.76^{+0.11}_{-0.09}$. The flux density measurements suggest a possible curve in the radio spectrum, which could be caused by synchrotron self-absorption or free-free absorption. However, more data points are necessary to be able to constrain any curvature.}
\end{figure*} 

PRIMER-COS 3866 has been identified as an X-ray source in the Chandra COSMOS Survey \citep{Elvis2009ApJS..184..158E} as CXOC100024.2+022510. A few years later, it was spectroscopically confirmed at $z=4.66$ by \cite{Civano2011ApJ...741...91C}. Based on recent JWST photometry, its SED is also in agreement with the source harboring an AGN \citep{Kocevski2024arXiv240403576K}. We confidently detect this source in the radio in MIGHTEE DR1 (S/N$\sim$8) and VLA-COSMOS (S/N$\sim$6). It is only tentatively detected with LOFAR (S/N$\sim$2). The radio images, optical overlay, and flux density measurements are shown in Fig.~\ref{fig:primer_cos_3866}. Its radio spectrum includes 3$\sigma$ upper limits on the flux density at 325 and 610 MHz obtained using the Giant Metrewave Radio Telescope (GMRT; \citealt{Swarup1991ASPC...19..376S}) with a local rms of 68 and 32 $\mu$Jy beam$^{-1}$, respectively (see \citealt{Tisanic2019A&A...621A.139T} for details). To constrain its spectral index we fit a simple power law\footnote{Defined as S$_{\nu}$ = c $\nu^{\alpha}$, with $\nu$ the frequency and $\alpha$ the spectral index.}, which is expected to be produced by non-thermal synchrotron emission. For the fitting process, we apply an invariant Markov Chain Monte Carlo (MCMC) method, utilizing the \textsc{emcee} python package \citep{emcee2013PASP..125..306F}. The best-fit parameters are determined by calculating the 50th percentile after 5,000 iterations, with the uncertainties estimated from the 16th and 84th percentiles. This yields a radio spectral index $\alpha=-0.76^{+0.11}_{-0.09}$, which is close in value to the typical extragalactic source radio spectral index of $\alpha=-0.7$ (e.g. \citealt{Hardcastle2016MNRAS.462.1910H, smolcic2017A&A...602A...6S}). As seen in Fig.~\ref{fig:primer_cos_3866} the radio spectrum of PRIMER-COS 3866 shows a possible curve, which could be caused by synchrotron self-absorption or free-free absorption. Similarly, previous work by \cite{Tisanic2020A&A...643A..51T} found that the average radio SED of AGN in COSMOS are best fit by a broken power law model, indicating that the SED shape is affected by ageing or absorption. However, in our case, more data points are necessary to test this hypothesis properly. 

Next, we determine the radio luminosity of PRIMER-COS 3866 using
\begin{equation}
    L_{R} = \frac{F_{\nu} \times 4 \pi D_L^2}{(1+z)^{1+\alpha}} \ \text{W Hz}^{-1},
\end{equation}
with $F_{\nu}$ the total flux density at observed frequency $\nu$, $D_L$ the luminosity distance, and $\alpha$ the measured radio spectral index. This yields a rest-frame radio luminosity of $L_{\text{3GHz}}$ of 3$\times$10$^{24}$ W Hz$^{-1}$ or 9$\times$10$^{40}$ erg s$^{-1}$.

Finally, we determine the radio-loudness of PRIMER-COS 3866, which we define as $R = f_{\text{5GHz}}/f_{4400\text{\AA}}$ with the flux values determined at rest-frame. To obtain $f_{4400\text{\AA}}$, we utilize the available JWST photometry\footnote{\url{https://dawn-cph.github.io/dja}} (in the filters F115W, F150W, F200W, F277W, F356W, F444W) and interpolate between these flux values in log space for both the wavelength and flux. Finally, we fold in a (1+$z$) correction to convert to rest-frame, which yields $f_{4400\text{\AA}} = 119\ \mu$Jy \footnote{\cite{Kocevski2024arXiv240403576K} extrapolate their best fit to the blue continuum to determine the rest-frame 1450\AA\ magnitude of PRIMER-COS 3866, which results in lower flux values due to their flat UV slope of $\beta_{\text{UV}} = -0.48$.}. Together with its radio flux density converted to 5 GHz rest-frame of $f_{\text{5GHz}}=60\ \mu$Jy, this results in a radio-loudness of $R\approx0.5$, which is far below the radio-loudness limit of $R=10$, classifying it as radio-quiet. This value is quite low compared to most radio detected high-$z$ AGN with typical measured radio-loudness values of $R\gtrsim1$ (e.g. \citealt{Banados2015ApJ...804..118B, Gloudemans2021A&A...656A.137G}). However, this could be due to the limited sensitivity of large sky radio surveys and rareness of known high-$z$ AGN. Compared with the radio-loudness distribution derived by \cite{Macfarlane2021MNRAS.506.5888M} for quasars from the Sloan Digital Sky Survey (SDSS; \citealt{York2000AJ....120.1579Y}) at low-$z$, PRIMER-COS 3866 falls in a region occupied by both AGN and star formation dominated sources. Note that this calculation does not take into account potential dust obscuration, which means that the radio-loudness value could be lower in reality. 

The question is whether this radio emission could also be generated from star formation. The observed radio spectral index does not distinguish between star formation and AGN activity (see e.g.~\citealt{Hardcastle2016MNRAS.462.1910H, An2021MNRAS.507.2643A}). One of the ways to unambiguously identify AGN activity from star formation is via the brightness temperature. The brightness temperature T$_b$ can be determined using 
\begin{equation}
    T_b = \Bigg(\frac{S_{\nu} [\text{Jy}]}{\theta_1\text{[mas]}\times\theta_2\text{[mas]}}\Bigg) \times \Bigg(\frac{1.22\times 10^{12}}{\nu [\text{GHz}]^2}\Bigg)\times (1+z) \,\text{K}
\end{equation}
with $S_{\nu}$ the total flux density, $\nu$ the observed frequency, and $\theta_1$ and $\theta_2$ the fitted minor and major axis of a 2D Gaussian (see e.g. \citealt{Morabito2022MNRAS.515.5758M}). Since PRIMER-COS 3866 is not resolved in the radio images, this measurements only provides a lower limit. The LOFAR flux measurements provides the most constraining $T_b$ measurement of $T_b \gtrsim 10^{3}$ K. This value is below the AGN limit of $T_b \sim 10^{6}$ K and therefore this does not prove the AGN nature of PRIMER-COS 3866. If the radio emission is generated purely by star formation, we expect an SFR of $\sim$820 M$_{\odot}$ yr$^{-1}$ from the observed radio luminosity - star formation rate correlation at 150 MHz \citep{Smith2021A&A...648A...6S}. Even though its radio luminosity could be explained by star formation, its high X-ray luminosity of L$_X \sim 5\times10^{44}$ erg s$^{-1}$ \citep{Kocevski2024arXiv240403576K} implies an extreme SFR of $\sim$10$^{5}$ M$_{\odot}$ yr$^{-1}$ (using \citealt{Mineo2014MNRAS.437.1698M}), making an AGN origin more plausible.  

PRIMER-COS 3866 does not have a JWST spectrum and therefore does not have an H$\alpha$ emission line detection or reliable black hole mass estimate. \cite{Kocevski2024arXiv240403576K} estimated it has a black hole mass to be $\sim8\times10^7$ M$_{\odot}$ from its X-ray luminosity and SED modeling, which is higher than most other spectroscopically confirmed AGN candidates (see Sect.~\ref{sec:discussion}). PRIMER-COS 3866 represents a special case within this sample of JWST AGN candidates, and therefore, its results do not directly apply to the (potentially) multiple populations considered in this work. 

\begin{figure*}
    \centering
    \includegraphics[width=0.85\linewidth]{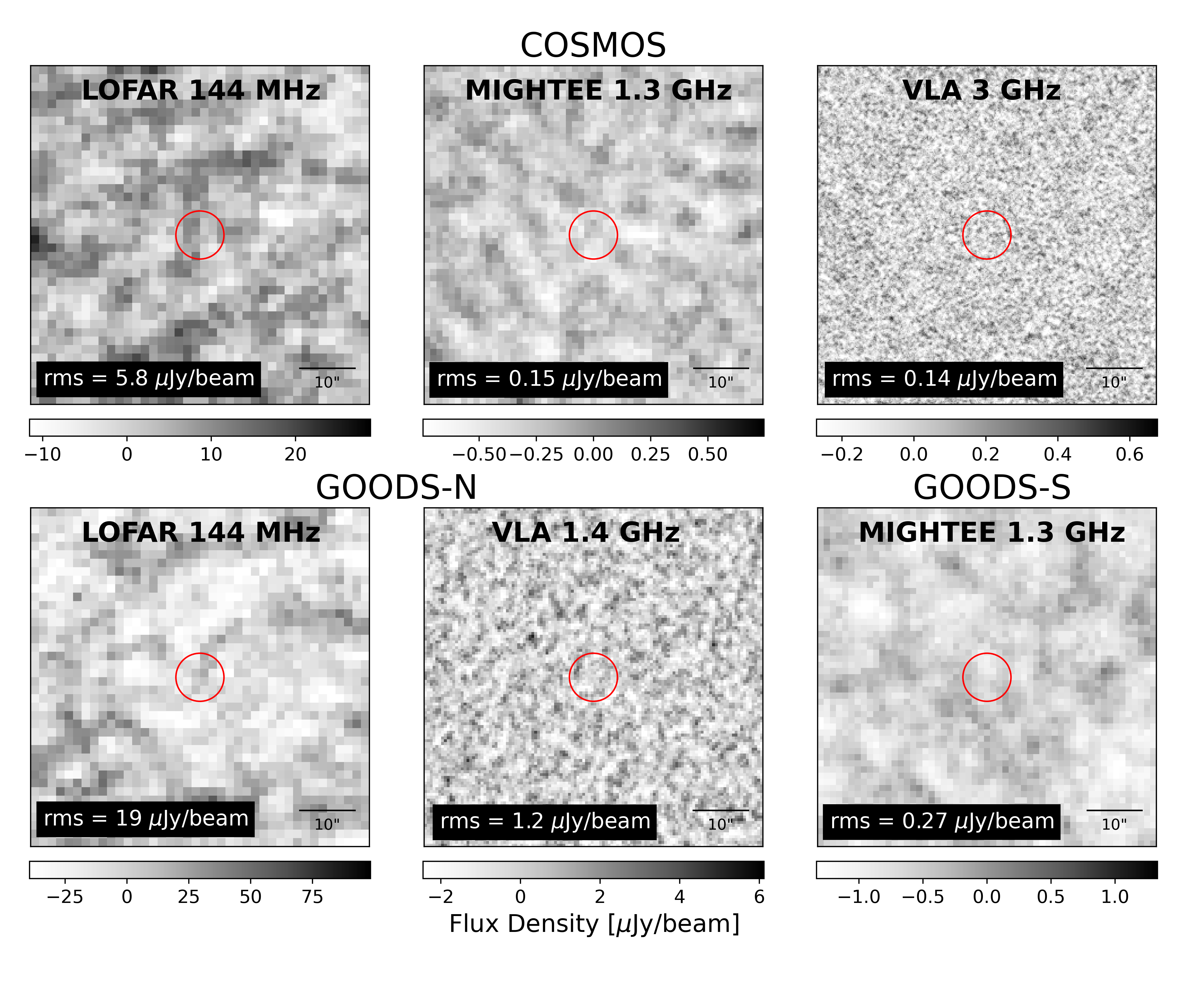}\vspace{-0.3cm}
    \caption{Stacked radio images of 593, 17, and 86 AGN candidates in COSMOS (top row), GOODS-N (bottom left and middle), and GOODS-S (bottom right), respectively. All stacks resulted in a non-detection with the measured rms shown in the lower left corner of each panel. Only the stack in GOODS-N consists of purely spectroscopically selected AGN candidates (see Table~\ref{tab:tab_lrds}).} 
    \label{fig:stacks}
\end{figure*}

\section{Radio stacking}
\label{sec:stacking}

To determine the average radio properties of the JWST discovered AGN candidate population, we perform radio stacking in each of the fields and surveys listed in Table~\ref{tab:tab_radio_obs}. We make 60\arcsec\ $\times$ 60\arcsec\ cutouts at all optical source positions and median stack them separately for each survey. We assume the sources are unresolved and obtain the median flux density from the peak pixels at the source location. None of the stacks result in a radio detection with all peak flux density measurements below $2\sigma$ significance. The 3$\sigma$ limits on the flux density are given in the final column of Table~\ref{tab:tab_radio_obs}. All stacks are shown in Fig.~\ref{fig:stacks}. 

The stacked radio images in GOODS-N, consisting of only spectroscopically confirmed AGN candidates, resulted in an average rms noise levels of $\sim$19 and 1.2 $\mu$Jy beam$^{-1}$ at 144 MHz and 1.4 GHz, respectively. This yields a 3$\sigma$ upper limit on the radio luminosity of  $L_{1.4\text{GHz}} = $ 8.6$\times$10$^{39}$ erg s$^{-1}$ (with $z_{\text{median}} = 5.2$). The stacks in the COSMOS field, consisting of photometrically selected AGN candidates, provide the deepest constraints with an rms of 0.15 and 0.14 $\mu$Jy beam$^{-1}$ at 1.3 and 3 GHz, respectively, which yield 3$\sigma$ upper limits on the radio luminosity of $L_{1.3\text{GHz}} \lesssim 1.3\times10^{39}$ erg s$^{-1}$ and $L_{3\text{GHz}} \lesssim 2.8\times10^{39}$ erg s$^{-1}$, respectively. However, since only 1 source (PRIMER-COS 3866; see Sect.~\ref{sec:primer_cos_3866_detection}) has been spectroscopically confirmed, these limits are uncertain.  

Furthermore, we determine an upper limit on the radio-loudness using the rest-frame optical flux at 4400\AA\ and limit on the radio flux density at 5 GHz rest-frame. 
Similar to our method in Sect.~\ref{sec:primer_cos_3866_detection}, we obtain $f_{4400\text{\AA}}$ values by interpolating (extrapolating for $z\gtrsim9$) in log space between the measured JWST flux values and correcting them to rest-frame by a factor (1+$z$). For the sample of \cite{Akins2024arXiv240610341A}, we use their provided flux measurements in filters F115W, F150W, F277W, F444W. For 4 sources in the sample of \cite{Matthee2024ApJ...963..129M} we instead extrapolate their derived $M_{1500\text{\AA}}$ value to 4400 \AA\ due to a lack of available photometry, assuming a K-correction of $-2$ and their published $\beta_{UV}$ values. In each field we determine the median $f_{4400\text{\AA}}$ value of all sources.
For the radio spectral index, we assume $\alpha=-0.7$. This yields a 3$\sigma$ upper limit on the radio-loudness of $R\lesssim18$ in GOODS-N (spectroscopically selected sample) and $R\lesssim2$ in COSMOS (photometrically selected sample). 
We note that the radio luminosity and radio-loudness measurements depend strongly on the assumed spectral index. For example, our radio-loudness limit in the COSMOS field changes from $R\lesssim1$ to $R\lesssim5$ when assuming $\alpha=0$ and $-2.5$, respectively. These radio-loudness limits in COSMOS are below the limit of $R=10$, however, these values are still similar to the radio-loudness found for quasars at $z>5$ (e.g. \citealt{Banados2015ApJ...804..118B, Gloudemans2021A&A...656A.137G}) and consistent with both AGN and star-formation dominated quasars from \cite{Macfarlane2021MNRAS.506.5888M}. Again, this radio-loudness calculation does not take into account potential dust-obscuration, therefore, providing an upper limit also in that sense.

 \begin{figure*}
    \centering
    \includegraphics[width=\textwidth]{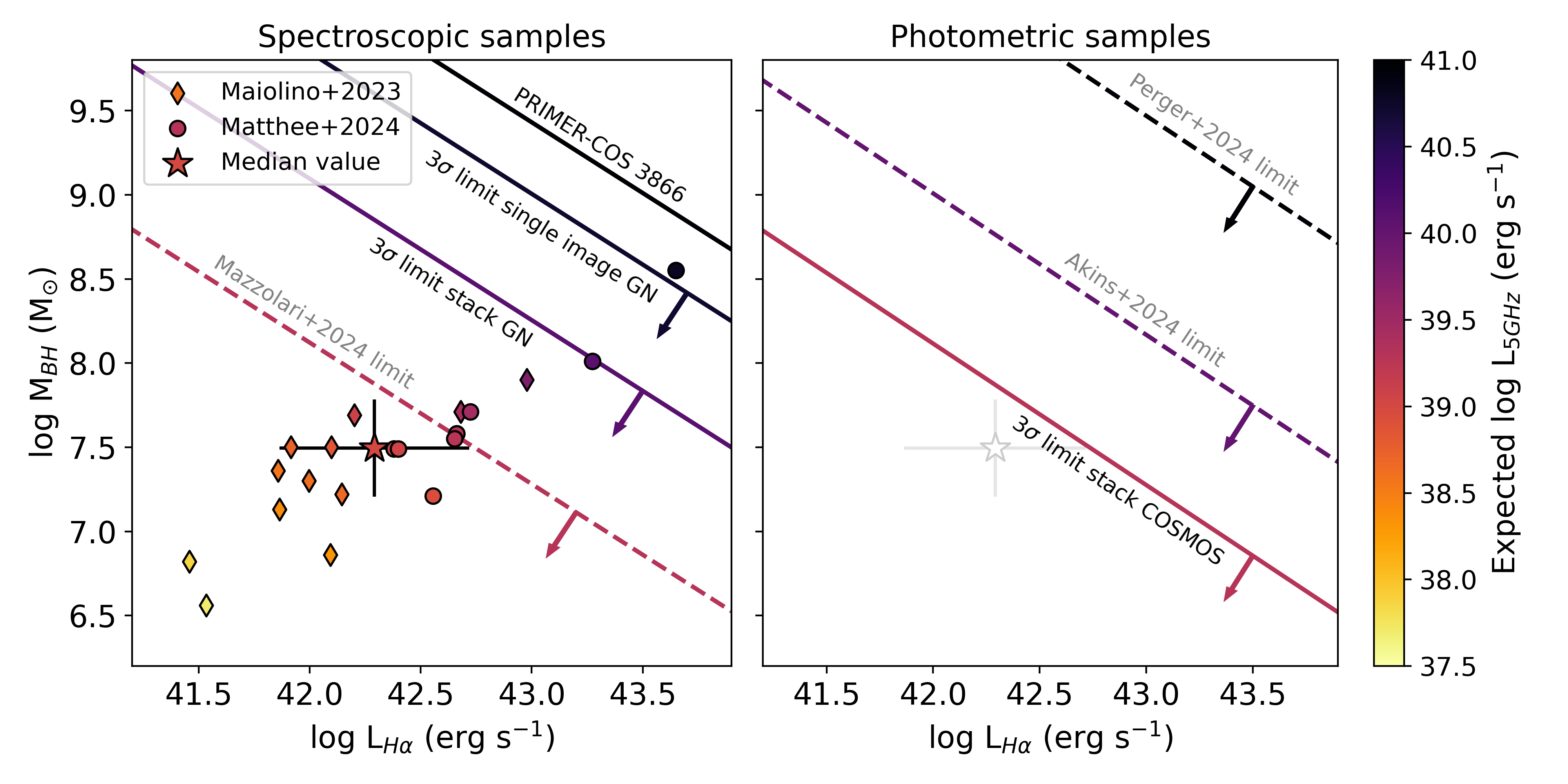}
    \caption{Black hole mass versus H$\alpha$ luminosity with predictions for the radio luminosity at 5 GHz rest-frame for spectroscopically selected AGN candidates (left) and photometrically selected AGN candidates (right). \textbf{Left:} Expected L$_{5\text{GHz}}$ for AGN candidates in GOODS-N and GOODS-S according to their BH mass and L$_{\text{H}\alpha}$. The 3$\sigma$ limits from our radio images are indicated for a single VLA image in GOODS-N, the VLA stack in GOODS-N. The black solid line gives the expected BH mass and H$\alpha$ luminosity values for PRIMER-COS 3866 as derived from its measured radio luminosity. The median value of the population in GOODS-N and GOODS-S is indicated by the star. The dashed line gives the 3$\sigma$ upper limit derived by \cite{Mazzolari2024arXiv241204224M}. \textbf{Right:} The 3$\sigma$ limits derived from the MIGHTEE stack in COSMOS using photometrically selected sources. The dashed lines show the 3$\sigma$ limits derived by \cite{Akins2024arXiv240610341A} and \cite{Perger2024arXiv241119518P}. The median value of the spectroscopic sample (gray star) is displayed for comparison. The median value is below all 3$\sigma$ limits derived in this work and literature. We conclude that not finding a radio detection is not (yet) in tension with their expected radio emission.} 
    \label{fig:expected_radio}
\end{figure*}

\section{Discussion}
\label{sec:discussion}

\subsection{Is a radio non-detection consistent with expectations?}
\label{subsec:expected_radio}

The stacking procedure presented above leads to a non-detection of the sources in all three fields. To estimate the radio luminosity that we would expect for the AGN candidates, we utilise the empirical correlations between radio ($L_R$) and X-ray luminosity ($L_X$) and X-ray and H$\alpha$ line luminosity ($L_{\text{H}\alpha}$) established for local AGN. There are various fundamental plane relations in the literature relating $L_R$, $L_X$, and the black hole mass $M_{\text{BH}}$ (e.g. \citealt{Merloni2003MNRAS.345.1057M, Saikia2015MNRAS.450.2317S, Bariuan2022MNRAS.513.4673B, Wang2024A&A...689A.327W}). In this work, we use the fundamental plane relation from \cite{Saikia2015MNRAS.450.2317S}, since their sample covers a wide range of BH masses and radio luminosities and gives the most conservative expected radio luminosity estimate for AGN with low BH masses ($\lesssim 10^8$ M$_{\odot}$).  

The observed correlation between $L_R$ and $L_X$ from \cite{Saikia2015MNRAS.450.2317S} is given by
\begin{equation}
\label{eq:radio_xray}
    \text{log} \ L_{15\text{GHz}} = 0.83 \ \text{log} \ L_X + 0.82 \ \text{log} \ M_{\text{BH}} -3.08 
\end{equation}
For consistency, we convert all radio luminosities to a 5 GHz rest-frame frequency using
\begin{equation}
\label{eq:radio_lum}
    L_{5\text{GHz}} = L_{\nu,\text{old}} \times \Big( \frac{5 \ \text{GHz}}{\nu_{\text{old}}} \Big)^{1+\alpha} \text{erg s}^{-1},
\end{equation}
assuming a spectral index of $\alpha = -0.7$. Furthermore, we adopt the correlation between $L_X$ and $L_{\text{H}\alpha}$ from \cite{Jin2012MNRAS.422.3268J}, which is given by
\begin{equation}
    \text{log} \ L_X  = 0.83 \ \text{log} \ L_{\text{H}\alpha} + 8.35 \ . 
\end{equation}
with a $\sim$0.3 dex scatter. This relation is obtained from a well-defined, predominantly radio-quiet, sample of 51 low-$z$ unobscured type I AGN. Combining these two equations gives a relation between $L_{\text{H}\alpha}$, $M_{\text{BH}}$, and $L_R$.  

We use the black hole masses and $L_{\text{H}\alpha}$\footnote{Three sources require two broad components to fit the H$\alpha$ line (see \citealt{Maiolino2023arXiv230801230M}). For these sources, we sum the fluxes of both components to obtain the H$\alpha$ luminosity.} of the sources in GOODS-N and GOODS-S and estimate their expected radio luminosity (see left panel of Fig.~\ref{fig:expected_radio}). The resulting predicted radio luminosities at 5 GHz rest-frame range between $5\times10^{37} - 6\times10^{40}$ erg s$^{-1}$ with a median value of $1\times10^{39}$ erg s$^{-1}$. We compare this to the 3$\sigma$ radio flux limits derived from single and stacked VLA images in GOODS-N, which provide the deepest radio limit from spectroscopically confirmed AGN candidates. As can be seen in Fig.~\ref{fig:expected_radio}, there is only 1 source (GOODS-N-9771; \citealt{Matthee2024ApJ...963..129M}) above the 3$\sigma$ limit of the single images, due to its high BH mass of 10$^{8.55}$ M$_{\odot}$. GOODS-N-9771 is however not detected in any of the radio images. The measured 3$\sigma$ limit of the stack in GOODS-N is higher than the median BH mass and H$\alpha$ luminosity value derived for the AGN candidates, and therefore their radio non-detection is as expected from these relations. We could not include the AGN candidates from \cite{Kocevski2024arXiv240403576K}, \cite{Akins2024arXiv240610341A}, and \cite{Kokorev2024ApJ...968...38K} in this figure, since these do not have spectral measurements. For PRIMER-COS 3866, we include expected possible values derived from its radio luminosity, which are shown by the black solid line.

Furthermore, we display the results for our photometrically selected samples separately in the right panel of Fig.~\ref{fig:expected_radio}. The MIGHTEE COSMOS and VLA-COSMOS data yield the deepest radio limits with 3$\sigma$ limits of 0.45 and 0.41 $\mu$Jy at 1.3 GHz and 3 GHz, respectively, or L$_{5\text{GHz}}<2.0\times10^{39}$ and 3.3$\times10^{39}$ erg s$^{-1}$ when assuming $\alpha=-0.7$ and $z_{\text{median}}=6.1$. This 3$\sigma$ limit on the MIGHTEE stack is only slightly higher than the expected radio luminosity derived from the median H$\alpha$ luminosity (of $1\times10^{39}$ erg s$^{-1}$). Directly from the limit on their X-ray luminosity of $L_X\sim10^{42-43}$ erg s$^{-1}$ (e.g. \citealt{Yue2024ApJ...974L..26Y, Maiolino2024arXiv240500504M, Inayoshi2024arXiv241203653I}), the relation in Eq.~\ref{eq:radio_xray} yields an expected $L_{5\text{GHz}}<10^{37-39}$ erg s$^{-1}$, which is also in line with a radio non-detection. 

We note that the expected radio luminosities vary when considering different fundamental plane relations. For example, the median expected radio luminosity of the spectroscopically selected AGN candidates changes from $L_{5\text{GHz}} = 1\times10^{39}$ erg s$^{-1}$ to $2\times10^{39}$ erg s$^{-1}$ for \cite{Merloni2003MNRAS.345.1057M}, $7\times10^{39}$ erg s$^{-1}$ for \cite{Bariuan2022MNRAS.513.4673B}, and $5\times10^{39}$ erg s$^{-1}$ for \cite{Wang2024A&A...689A.327W}. These relations have been derived from different AGN samples and mainly diverge at low BH mass ($\lesssim10^8$ M$_\odot$). The relations from \cite{Bariuan2022MNRAS.513.4673B} and \cite{Wang2024A&A...689A.327W} suggest the radio non-detection of our sample is not as expected. However, the relation from \cite{Saikia2015MNRAS.450.2317S} has been derived from AGN with a wider range of radio luminosities down to $10^{36}$ erg s$^{-1}$ and BH masses down to $10^7$ M$_\odot$, and provides the most conservative estimate on the expected radio luminosity. We therefore conclude that the current limits on the radio stacks are not constraining enough to claim these JWST discovered AGN candidates are radio-weaker than expected, and deeper radio surveys are necessary to obtain a detection or a meaningful upper limit.

In addition to PRIMER-COS 3866, there is another X-ray detected JWST discovered AGN candidate, JADES 21925 at $z=3.1$ in GOODS-S (see \citealt{Kocevski2024arXiv240403576K}). This source is however not detected in MIGHTEE with a 3$\sigma$ flux density upper limit of 5.7 $\mu$Jy beam$^{-1}$ and radio luminosity limit of 6$\times$10$^{39}$ erg s$^{-1}$ at 5 GHz rest-frame. Comparing this to its rest-frame optical magnitude at 4400 \AA\ suggests a radio-loudness limit of $R<11$. From its X-ray luminosity of $\log (L_{2-10 \text{keV}}$/ erg s$^{-1}$) = 43.73$\pm$0.06 and black hole mass of $\sim$8.5$\times$10$^{6}$ M$_{\odot}$, we would expect a radio luminosity of $L_{5\text{GHz}} \approx 6\times10^{38}$ erg s$^{-1}$ (see eq.~\ref{eq:radio_xray}), which is $\sim$10$\times$ lower than the derived upper limit. Therefore the radio non-detection is again as expected.  

\subsection{Literature comparison}

This study was conducted in parallel to the analysis of \cite{Mazzolari2024arXiv241204224M}. In their study of the GOODS-N field, they find a 3$\sigma$ upper limit on the radio luminosity of L$_{5\text{GHz}} < 2\times10^{39}$ erg s$^{-1}$, which is similar to our limit found in the COSMOS field for photometrically selected AGN candidates. By considering local scaling relations, they suggest that the JWST detected broad line AGN might show radio-weak behavior. Furthermore, \cite{Akins2024arXiv240610341A} find their radio-loudness limit (using $R_{1.4} = L_{\nu,1.4\text{GHz}}/L_{\nu,4400}$) to be below the limit with a radio luminosity of $L_{\nu,1.4\text{GHz}} < 5.2\times10^{23}$ W Hz$^{-1}$. Finally, \cite{Perger2024arXiv241119518P} also find a non-detection of LRDs down to a 3$\sigma$ noise levels of 11 $\mu$Jy beam$^{-1}$ and conclude that LRDs host weaker (sub-dominant) or no radio AGN. To compare these results to our own, we convert these limits to 5 GHz rest-frame and show them in Fig.~\ref{fig:expected_radio}. All limits are still above the median expected radio luminosity. Therefore, these relations suggest that none of these radio measurements are deep enough to detect the median population.

While these JWST discovered AGN candidates might indeed be radio-weak, this work suggests that current radio observations are not constraining enough to confirm the possible radio-weakness of the JWST discovered AGN candidates and deeper radio observations are necessary. We furthermore speculate that any radio signal in the stack could be diminished by the inclusion of very faint sources and contaminants in the sample, which again highlights the need for further investigation.

\subsection{Future surveys}
\label{subsec:future_surveys}

This work demonstrates that deeper radio observations are necessary to detect the general population of JWST discovered AGN candidates (including LRDs). With our calculation in Sect.~\ref{subsec:expected_radio}, we expect these sources to have a median radio luminosity of $1\times10^{39}$ erg s$^{-1}$ at 5 GHz (equal to $\sim$0.13 $\mu$Jy at $z\sim5$). This sensitivity regime will be accessible by future radio telescopes such as the Square Kilometer Array (SKA; \citealt{Dewdney2009IEEEP..97.1482D}) and the Next Generation Very Large Array (ngVLA; \citealt{McKinnon2019BAAS...51g..81M}). 

The SKA is expected to have a weighted continuum sensitivity of 1.2-17 $\mu$Jy beam$^{-1}$ hr$^{-1}$ at 1.3 GHz in the AA$^{*}$ configuration\footnote{\url{https://sensitivity-calculator.skao.int/}}. This large range is caused by the choice of image weighting (e.g. natural versus uniform) and the sensitivity will depend strongly on the observing band, subarray configuration, and sky position. However, this example demonstrates that a sensitivity of $\sim$1 $\mu$Jy beam$^{-1}$ can easily be reached, which would allow for the $>3\sigma$ detection of $\gtrsim80$ stacked AGN candidates from our predictions in Sect.~\ref{subsec:expected_radio}. 

The ngVLA is expected to reach a continuum rms of 0.24 $\mu$Jy beam$^{-1}$ hr$^{-1}$ at 2.4 GHz using natural image weights and all baselines\footnote{See \url{https://ngvla.nrao.edu/page/performance}}. At this sensitivity we can expect to detect individual sources as well as obtain a solid 3$\sigma$ detection of the general population by stacking $>10$ sources. Constraining their brightness temperature, radio-loudness, and radio spectral index will enable to break the SF/AGN degeneracy. 

Our understanding will furthermore be improved by the discovery of larger samples of AGN candidates including the spectroscopic confirmation of candidate samples (e.g. \citealt{Kocevski2024arXiv240403576K, Akins2024arXiv240610341A, Kokorev2024ApJ...968...38K}). Ultimately, a combination of spectroscopic follow-up (e.g. large grism surveys such as \citealt{Kakiichi2024jwst.prop.5893K}) and multi-wavelength characterization (including X-ray, IR, and radio) will be necessary to determine the true nature of these compact red sources. Another future prospect is using strong gravitational lensing to obtain more radio detections (e.g. \citealt{McCarty2024arXiv241201746M}).

\section{Summary}

We studied the radio properties of three samples of JWST discovered AGN candidates in COSMOS, GOODS-N, and GOODS-S using LOFAR, MeerKAT, and VLA in the frequency range of $0.144-3$ GHz. Our main findings are as follows:
\begin{itemize}
\setlength\itemsep{0.1em}

    \item The spectroscopically confirmed AGN candidate PRIMER-COS 3866 at $z=4.66$ is the only source that is detected in our radio surveys. Its radio spectrum yields a spectral index of $\alpha=-0.76^{+0.11}_{-0.09}$. The spectrum shows a possible curvature caused by absorption, however, more data points are necessary to constrain this. Its radio-loudness of $R\approx0.5$ classifies it as radio-quiet, but is still consistent with other known AGN. The derived lower limit on the brightness temperature of $T_b \gtrsim 10^{3}$ K is too low to confirm its AGN nature.   
    
    \item None of our stacks resulted in a radio detection with peak flux density measurements below 2$\sigma$ significance. The 3$\sigma$ upper limits on the radio-loudness constrain the JWST discovered AGN candidate population to $R\lesssim18$ in GOODS-N (spectroscopically confirmed sample) and $R\lesssim2$ in COSMOS (photometrically selected sample). However, these constraints still depend on their radio spectral indices and true redshifts in the case of the photometrically selected sources.
    
    \item The radio non-detections are still consistent with expectations from empirical $L_X - L_{\text{H}\alpha}$ and $L_X - L_R$ correlations. From their BH masses and H$\alpha$ luminosities, we determined an expected radio luminosity of $L_{5\text{GHz}}\sim$10$^{39}$ erg s$^{-1}$ of the population if these empirical relations hold. Note, however, that previous X-ray observations suggest that these are not regular type I AGN. We conclude that our current constraints from radio observations are not deep enough to determine if these JWST discovered AGN candidates are radio-weak.
    
    \item Future radio telescopes, such as SKA and ngVLA, are expected to be able to reach these sensitivity limits easily within a few hours of observations. Especially, a measurement of the temperature brightness of this population will be crucial for confirming their AGN nature. 
    
\end{itemize}

\section*{Acknowledgments}

K.J.D. acknowledges funding from the STFC through an Ernest Rutherford Fellowship (grant number ST/W003120/1). E.P.F. is supported by the international Gemini Observatory, a program of NSF NOIRLab, which is managed by the Association of Universities for Research in Astronomy (AURA) under a cooperative agreement with the U.S. National Science Foundation, on behalf of the Gemini partnership of Argentina, Brazil, Canada, Chile, the Republic of Korea, and the United States of America. K. I. acknowledges support from the National Natural Science Foundation of China (12073003, 12003003, 11721303, 11991052, 11950410493), and the China Manned Space Project (CMSCSST-2021-A04 and CMS-CSST-2021-A06).

This paper is based on data obtained with the
LOFAR telescope (LOFAR-ERIC). LOFAR (van Haarlem et al.
2013) is the Low Frequency Array designed and constructed by ASTRON. It has observing, data processing, and data storage
facilities in several countries, that are owned by various parties (each with their own funding sources), and that are collectively
operated by the LOFAR European Research Infrastructure Consortium (LOFAR-ERIC) under a joint scientific policy. The
LOFAR-ERIC resources have benefited from the following recent major funding sources: CNRS-INSU, Observatoire de Paris
and Universit\'e d’Orl\'eans, France; BMBF, MIWF-NRW, MPG, Germany; Science Foundation Ireland (SFI), Department of
Business, Enterprise and Innovation (DBEI), Ireland; NWO, The Netherlands; The Science and Technology Facilities Council,
UK; Ministry of Science and Higher Education, Poland.

The research leading to these results has received funding from the European Union’s Horizon 2020 research and innovation programme under grant agreement No 101004719 [ORP].


\appendix

\section{False radio detections}
\label{appendix:radio_association}

\begin{figure*}
    \centering
    \includegraphics[width=\textwidth]{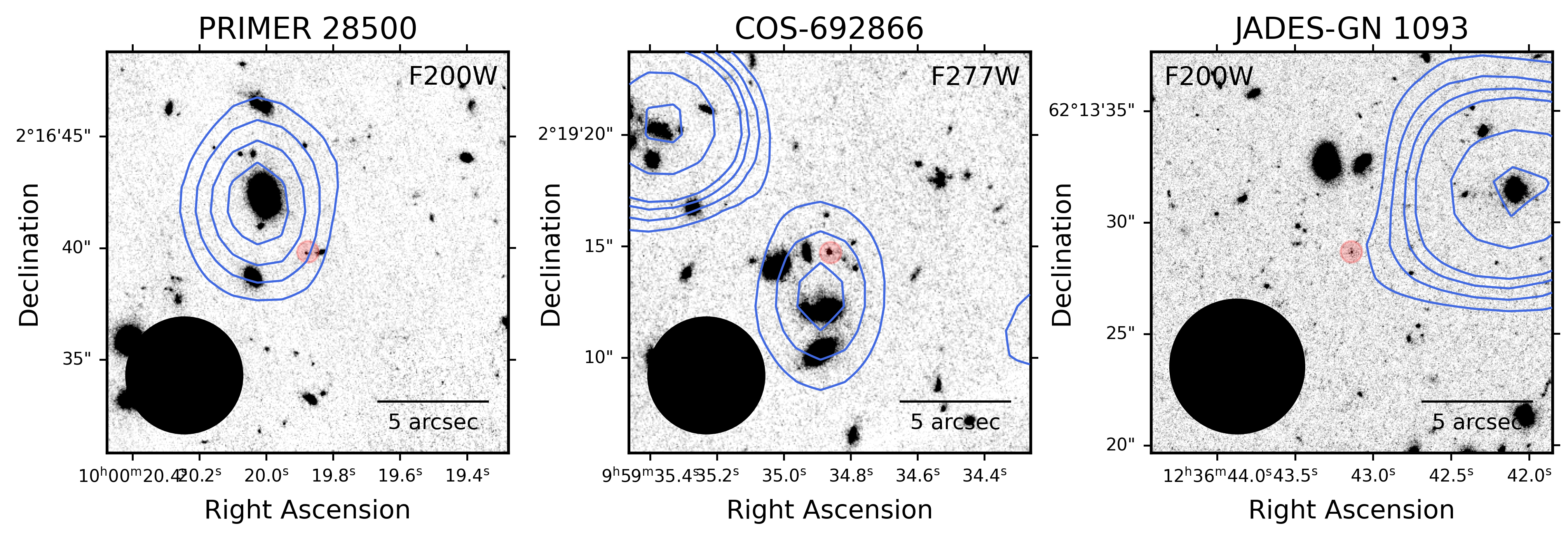}
    \caption{Three random examples of JWST discovered AGN candidates with wrongly associated potential radio detections from MIGHTEE (left and middle) and LOFAR (right panel). The radio contours are drawn at [$-1$, 0.5, 1, 1.5, 2, 4, 6, 8]$\ \times\ 3\sigma_{\text{rms}}$ and the beam size is shown in the bottom left. The red circles highlight the AGN candidates. }
    \label{fig:examples_wrong_association}

\end{figure*}

Figure \ref{fig:examples_wrong_association} shows a few examples of false radio detections of our AGN candidates, where the radio source, in reality, is associated with a bright galaxy in the vicinity. Sometimes these bright galaxies are within the beam causing the confusion, but often it's diffuse extended emission that is causing the false radio detection. Especially in the case of low-resolution radio observations and highly crowded fields, it is crucial to visually inspect any radio detection or perform likelihood association calculations.  

\bibliography{main}{}
\bibliographystyle{aasjournal}

\end{document}